\documentclass[preprint,nofootinbib,aps,superscriptaddress,eqsecnum]{revtex4-1} 
 \pdfoutput=1
 \usepackage{graphicx}
 \usepackage{amsmath}
\usepackage{amsfonts}
\usepackage{amssymb}
\usepackage{hyperref}
\usepackage{gensymb}
\usepackage{color}
\usepackage{soul}
\usepackage{caption}
\usepackage{subcaption}
\def\bea{\begin{eqnarray}}
\def\eea{\end{eqnarray}}
 \def\be{\begin{equation}}
\def\ee{\end{equation}}

\begin{document}

\title{Leptogenesis and {\it eV} scale sterile neutrino }

 \author{Srubabati Goswami}
\email{sruba@prl.res.in}
\affiliation{Theoretical Physics Division, Physical Research Laboratory, Ahmedabad - 380009, India}

\author{Vishnudath K. N.}
\email{vishnudathkn@imsc.res.in}
\affiliation{The Institute of Mathematical Sciences,
C.I.T Campus, Taramani, Chennai 600 113, India}

\author{Ananya Mukherjee}
\email{ananyatezpur@gmail.com}
\affiliation{Department of Physics, University of Calcutta, 92 Acharya Prafulla Chandra Road, Kolkata 700 009, India}

\author{Nimmala Narendra}
\email{nnarendra@prl.res.in}
\affiliation{Theoretical Physics Division, Physical Research Laboratory, Ahmedabad - 380009, India}

\begin{abstract}

We consider the minimal extended seesaw model which can accommodate an eV scale sterile neutrino. The scenario also includes three heavy right handed neutrinos in addition to the light sterile neutrino. In this model, the active-sterile mixing act as non-unitary parameters. If the values of these mixing angles are of $\mathcal{O}(0.1)$, the model introduces deviation of the PMNS matrix from unitarity to this order. We find that the oscillation data from various experiments imposes an upper bound on the lightest heavy neutrino mass scale as $\sim 10^{11}$ GeV in the context of this model. We study {\it vanilla} leptogenesis in this scheme, where the decay of the heavy right handed neutrinos in the early universe can give rise to the observed baryon asymmetry. Here, even though the eV scale sterile neutrino does not participate directly in leptogenesis, its effect is manifested through the non-unitary effects. We find that the parameter space that can give rise to successful leptogenesis is constrained by the bounds on the active-sterile mixing as obtained from the global analysis. 

\end{abstract}

\pacs{}
\maketitle

\section{Introduction}

Many short baseline experiments suggest the existence of at least one light sterile neutrino of mass in the eV scale. The first hint for this came from the $\overline{\nu_\mu} \rightarrow \overline{\nu_e}$ searches in the LSND experiment~\cite{Aguilar:2001ty}. Recently, the MiniBooNE experiment also confirmed this from the excess of the electron-like events~\cite{Aguilar-Arevalo:2018gpe}. Earlier, the reactor~\cite{Declais:1994su,Mention:2011rk} and the Gallium~\cite{Anselmann:1994ar,Hampel:1997fc,Giunti:2010zu,Abdurashitov:1996dp} anomalies also indicated the presence of an extra sterile neutrino that mixes with the three active flavor states of the Standard Model~(SM). For a recent analysis, see \cite{Berryman:2021yan}. The 3+1 picture, which was first introduced in \cite{Goswami:1995yq}, is the minimal scheme that can explain these anomalies. However, no signal of a fourth sterile neutrino has been reported in the disappearance experiments using neutrinos from reactors and accelerators~\cite{Giunti:2019aiy}. The results of the fit of short-baseline neutrino oscillation data in the framework of 3+1 active-sterile neutrino mixing are given in~\cite{Gariazzo:2017fdh,Dentler:2018sju}. See \cite{Dasgupta:2021ies} for a recent review on the status and phenomenology of eV scale sterile neutrino \footnote{Note that an eV scale sterile neutrino is disfavored by cosmology in general.  The secret interaction model of sterile neutrinos  proposed in \cite{Dasgupta:2013zpn} to ameliorate the situation was also disfavored later~\cite{Forastieri:2017oma,Song:2018zyl,Chu:2018gxk}. However, the recent analysis done in \cite{Archidiacono:2020yey} with a pseudoscalar interaction admits a sterile neutrino of mass $\sim 1$ eV.}\,. Also, note that the recent results from the analysis of the three years of data from MicroBooNE showed no excess of electrons. However,  this does not yet conclusively prove that sterile neutrino solution to MiniBooNE data is ruled out~\cite{Arguelles:2021meu}.  Also, the analysis done in reference \cite{Denton:2021czb} shows that the oscillations to sterile neutrino is still possible with $\Delta m_{14}^2 \sim 1.5 ~\textrm{eV}^2$ and $\textrm{sin}^2\theta_{14} \gtrsim 0.1$ for the intrinsic electron neutrinos in the beam.

Theoretical models that explain non-zero active neutrino masses and can simultaneously incorporate an eV scale sterile neutrino have been proposed, for instance, in \cite{Chun:1995js,Chun:1995bb,Barger:2003xm,Mohapatra:2001ns}. An elegant scheme to accommodate a light sterile neutrino within the type-I seesaw framework is the minimal extended seesaw (MES) discussed in \cite{Barry:2011wb}. In the MES scheme, the SM is extended by three heavy Majorana right handed neutrinos (RHNs), $\nu_R$, whose masses are $\sim \mathcal{O}(10^{12})$~ GeV and another gauge singlet fermion, $\nu_s$. Among the three RHNs, two are responsible for making two SM neutrinos massive whereas the third one gives mass to the light sterile neutrino. In this model, the mixing between the active neutrinos and the light sterile neutrino act as the non-unitary parameters characterizing the deviation from unitarity of the PMNS matrix. Several works have attempted to implement this scheme in the context of discrete flavour symmetric groups such as $A_4$~\cite{Zhang:2011vh,Das:2018qyt,Krishnan:2020xeq}. 

In addition to explaining the non-zero neutrino masses and mixing, the seesaw mechanism can also address the issue of the baryon asymmetry of the universe to which the SM does not have an answer~\cite{Fukugita:1986hr}. The comoving baryon asymmetry of the universe is given as,
\be Y_B = \frac{(n_B - n_{\bar{B}})}{s} ,\ee where $n_B$ and $n_{\bar{B}}$ are the number densities of baryons and antibaryons respectively and $s$ is the entropy density. The combined analysis of the data
from measurements of cosmic microwave background and large scale structure indicates a $3\sigma$ range for the baryon
asymmetry of the universe as~\cite{Aghanim:2018eyx},
\be Y_B = (8.52 - 8.93)\times10^{-11} .\ee
In the type-I seesaw model, the out of equilibrium decay of the heavy Majorana RHNs in the early universe can generate a lepton asymmetry, which in turn can be converted into a baryon asymmetry via the non-perturbative sphaleron process \cite{Rubakov:1996vz}.

We study leptogenesis in the context of the MES model in this work. Specifically, we ask the question if the light sterile neutrino plays any role in leptogenesis. On a first look, it might appear to one as if the light sterile neutrino plays absolutely no role in leptogenesis. But on a closer analysis, one can find that the large mixing between the eV scale sterile and active neutrinos can have an impact on leptogenesis. In fact, the standard loop diagrams for the leptogenesis is generated by the dimension-5 operator and the inclusion of the non-unitarity effects will correspond to the inclusion of the contribution due to the dimension-6 operator. This has been studied at the operator level in the context of a three family low scale seesaw in reference \cite{Antusch:2009gn}. One can also study the non-unitarity effects in leptogenesis by using the Casas-Ibarra (CI) parametrization~\cite{Casas:2001sr} for the neutrino Yukawa coupling, $y_\nu$ (or equivalently, the neutrino Dirac mass term, $M_D$). The expression for the CP asymmetry in leptogenesis depends on $y_\nu$ and in CI parametrization, $y_\nu$ can be expressed in terms of $U_{PMNS}$ matrix, the light and heavy neutrino masses and a complex orthogonal matrix, $R$. Thus, the effect of non-unitarity can easily be incorporated through the dependence of $y_\nu$ on $U_{PMNS}$. The author of reference \cite{Rodejohann:2009cq} has studied the effect of non-unitarity in leptogenesis following this approach in the context of a variant of a type-I seesaw. In that model, the sources of non-unitarity and active light neutrino mass generation were decoupled in the sense that the non-unitarity was due to the mixing of active neutrinos with neutral fermions which were different from the ones that were responsible for light neutrino mass generations. In this work, we study the effects of an eV scale sterile neutrino with large active-sterile mixing to explain the LSND-MiniBooNE anomaly on leptogenesis. In our analysis, we find that the contribution to the CP asymmetry from the non-unitary part is comparable to the one due to the unitary part, making the inclusion of non-unitary effects important. The parameter space that can give rise to successful leptogenesis gets constrained by the bounds on the active-sterile mixing as obtained from the global analysis and thereby manifesting the effect of the light sterile neutrino.

The rest of this paper is organized as follows : In section~\ref{sec:MES}, the MES model is briefly reviewed and the CI parametrization for the neutrino Dirac Yukawa couplings is discussed in section~\ref{sec:YukawaCI}.  Basics of leptogenesis and the relevant working formulae are given in section~\ref{sec:leptogenesis}. The results of our analysis are discussed in section~\ref{sec:analysis} and finally, we conclude in section~\ref{sec:conclusion}.

\section{The minimal extended seesaw}\label{sec:MES}
In the MES model~ \cite{Barry:2011wb}, one adds three RHNs $\nu_R$ and one sterile neutrino $\nu_s$ to the SM particle content. The part of the Lagrangian relevant for neutrino mass generation is,
\be\label{eq:totalmm1}
- \mathcal{L}_Y =  y_\nu \bar l_L \tilde{H} \nu_R  + \bar \nu_R^c M_S
 \nu_s + \frac{1}{2} \bar \nu_R^c M_R \nu_R+ h.c.,
\ee
where $l_L$ is the lepton doublet and $H$ is the SM Higgs doublet with $\tilde{H} = i\sigma_2 H^*$. Note that in the above equation, the generation indices are suppressed and $y_\nu$ is the $3\times 3$ neutrino Yukawa coupling matrix and $M_S$ and $M_R$ are $3\times 1$ and $3 \times 3$ matrices respectively. Without lose of generality, we work in a basis in which $M_R$ is diagonal and real. We also take the charged lepton mass matrix to be diagonal. Once the electroweak symmetry is spontaneously broken, the Lagrangian in Eq.~(\ref{eq:totalmm1}) becomes,
\be\label{eq:totalmm}
 \mathcal{L}_\nu = \bar \nu_L M_D \nu_R  + \bar \nu_R^c M_S
 \nu_s + \frac{1}{2} \bar \nu_R^c M_R \nu_R+ h.c.,
\ee
where, $M_D = y_\nu v/\sqrt{2}$ and $v = 246$ GeV is the Higgs vacuum expectation value. The Lagrangian in Eq.~(\ref{eq:totalmm}) leads to the following $7 \times 7$ neutrino mass matrix in the ($\nu_L, \nu_s^c, \nu_R^c$) basis:
\be\label{M7}
 M_\nu^{7 \times 7} = \left(\begin{array}{ccc}
       0 & 0 & M_D \\
       0 & 0 & M_S^T\\
       M_D^T & M_S & M_R
       \end{array}\right). 
\ee
Assuming the mass terms to have a hierarchy as, $M_R >> M_S > M_D$, the RHNs that are much heavier compared to $v$ can be integrated out first. This results in the effective $4 \times 4$ light neutrino mass matrix in the ($\nu_L , \nu_s^c $) basis which is given as,
\be\label{eq:M4}
       M_\nu^{4 \times 4} = - \left(\begin{array}{cc}
       M_D M_R^{-1}M_D^T & M_D M_R^{-1} M_S^{T} \\
       M_S (M_R^{-1})^T M_D^T & M_S M_R^{-1}M_S^T 
       \end{array}\right).                 
\ee
This is a minimal extension of the type-I seesaw in the sense that
only one extra sterile field is added to the standard type-I seesaw scenario and the mass of this additional sterile field is also suppressed by $M_R$ along with that of the three active neutrinos. Since $M_\nu^{7 \times 7}$ has rank 6 and subsequently $M_\nu^{4 \times 4}$ has rank 3, the lightest neutrino state becomes massless. Now, since $M_S > M_D$, one may further integrate out the eV scale sterile state $\nu_s$ of mass,
\be\label{eq:sterilemass}
 m_4 \simeq M_S M_R^{-1}M_S^T,
\ee
from Eq.~(\ref {eq:M4}) to get the $3\times 3$ active light neutrino mass matrix as,
\be\label{eq:M3}
 M_\nu^{3\times 3} \simeq  M_D M_R^{-1}M_S^T (M_S M_R^{-1} M_S^T)^{-1} M_S M_R^{-1} M_D^T - M_D M_R^{-1} M_D^T.
\end{equation}
It is worth mentioning that the RHS of Eq.~(\ref{eq:M3}) remains non-vanishing since $M_S$ is a row vector and not a square matrix. In the standard picture with three active light Majorana neutrino mixing, the  relationship  between  the  flavor  and  mass  states is described by a $3\times 3$ unitary matrix, $U_\nu$, which  can  be  parameterized  in  terms  of three  mixing  angles ($\theta_{12}, \theta_{23}$ and $\theta_{13}$), one CP-violating phase ($\delta$) and two Majorana phases $(\alpha,\beta)$ (In our case, the lightest active neutrino is massless and this implies that $\beta = -\alpha$). Adding a sterile state expands the mixing  matrix  to  $4 \times 4$,  in  which  the  added  degrees  of  freedom  can  be  parameterized  by  introducing three new rotation angles ($\theta_{14},\theta_{24}, \,\,\text{and}\,\,\theta_{34}$), and two new  oscillation-accessible CP-violating  phases, $\delta_{14} \,\,\text{and}\,\, \delta_{24}$. In fact, in the above step of integrating out the eV scale sterile neutrino, the mass matrix in Eq.~(\ref{eq:M4}) can be diagonalised by the $4 \times 4$ unitary matrix that is given as (since we are neglecting the non-unitarity due to $\nu_R$, which goes as $M_D^2/M_R^2 \sim 10^{-20}$ taking $M_D \sim 100$ GeV and $M_R \sim 10^{12}$ GeV), 
\be\label{eq:u44}
U \simeq \left(\begin{array}{cc}
       (1-\frac{1}{2}V V^\dagger) U_\nu &  V \\
       -V ^\dagger U_\nu & 1-\frac{1}{2}V ^\dagger V 
       \end{array}\right).
\ee
In this equation, the three-component column vector $V$ is given by, 
\be\label{Req}
  V = M_D M_R^{-1}M_S^T (M_S M_R^{-1}M_S^T)^{-1} \equiv (U_{e4}, U_{\mu 4}, U_{\tau 4})^T,
\ee
and it corresponds to the active-sterile mixing, which is responsible for the non-unitarity of the PMNS matrix,
\be \label{eq:u33}
U_\text{PMNS} = (1-\frac{1}{2} V V^\dagger ) U_\nu.
\ee
Here, $U_\nu$ is the $3\times 3$ unitary PMNS matrix. Note that $V$ is suppressed by $\mathcal{O}(M_D / M_S)$ and hence the deviation of $U_\text{PMNS}$ from unitarity, i.e.~$-\frac{1}{2}U_LV V^\dagger U_\nu$, is  $\sim \mathcal{O}(M_D^2 / M_S^2)$.
\section{Casas-Ibarra parametrization for the Yukawa couplings} \label{sec:YukawaCI}

Using the CI parametrization \cite{Casas:2001sr}, one can express the Dirac mass matrix in terms of the $U_{PMNS}$ matrix, the light and heavy neutrino masses and a complex orthogonal matrix, $R$. In this section, we derive the CI parametrization for the Dirac mass matrix $M_D$ in the MES model. To do this, note that the light neutrino mass matrix in Eq.~(\ref{eq:M3}) can be written as,
\be
 M_\nu^{3\times 3} \simeq  M_D ( M_R^{-1}M_S^T (M_S M_R^{-1} M_S^T)^{-1} M_S M_R^{-1}  -  M_R^{-1} ) M_D^T = M_D A M_D^T,
\end{equation}
where we have denoted,
\be \label{A-def}
A= M_R^{-1}M_S^T (M_S M_R^{-1} M_S^T)^{-1} M_S M_R^{-1}  -  M_R^{-1},
\ee
which is a $3\times 3$ symmetric matrix. Now, $M_\nu^{3\times 3}$ can be diagonalized as,
\be \label{CIeq1} U_{PMNS}^T ( M_\nu^{3\times 3} ) U_{PMNS} \equiv U_{PMNS}^T (M_D A M_D^T) U_{PMNS} = D_m, \ee
where,
\be D_m = \textrm{diag} (m_1, m_2, m_3),  \ee
and $m_{1,2,3}$ are the light neutrino masses. Writing Eq.~(\ref{CIeq1}) as,
\be D_m = U_{PMNS}^T M_D \sqrt{A} \sqrt{A} M_D^T U_{PMNS}, \ee
and multiplying the left and right of this equation by $\sqrt{D_m^{-1}}$, we get,
\be \begin{split} I  = & \sqrt{D_m^{-1}} U_{PMNS}^T M_D \sqrt{A} \sqrt{A} M_D^T U_{PMNS}  \sqrt{D_m^{-1}} \\ = &  (\sqrt{A} M_D^T U_{PMNS}  \sqrt{D_m^{-1}})^T (\sqrt{A} M_D^T U_{PMNS}  \sqrt{D_m^{-1}}) \end{split}. \ee
Thus,
\be R = \sqrt{A} M_D^T U_{PMNS}  \sqrt{D_m^{-1}} \ee
is a $3\times 3$ orthogonal matrix. The above equation can be inverted to write $M_D$ as,
\begin{equation}\label{eq:diracCI}
M_D^T = (\sqrt{A})^{-1} R \sqrt{D_m}U_{PMNS}^{-1},  \,\,\,\,\text{or}\,\,\,\, M_D = U_{PMNS}^* \sqrt{D_m} R^T (\sqrt{A})^{-1}.
\end{equation}
This is the CI parametrization for $M_D$ in the MES model. Note that in the above equation, $A$ is given by Eq.~(\ref{A-def}) and $R$ is a general $3 \times 3$ complex orthogonal matrix. It is evident from the above expression that the scale of $M_D$ is guided by the scales of $M_S$ and $M_R$. Also, taking $U_{PMNS} =  (1-\frac{1}{2} V V^\dagger ) U_\nu $ in Eq.~(\ref{eq:diracCI}) will amount to incorporating the non-unitary effects (and hence both the the dimension-5 and dimension-6 contributions to leptogenesis) whereas taking $U_{PMNS} = U_\nu$ (an artificial case, where only the dimension-5 contribution is taken by putting $V=0$ by hand) will switch off the non-unitary effects.

\section{Baryogenesis through leptogenesis}\label{sec:leptogenesis}

It is well known that an out-of-equilibrium CP violating decay of the RHNs in the early universe can produce a lepton asymmetry which in turn can be converted into the baryon asymmetry dynamically (for details, see \cite{Buchmuller:2004nz,Buchmuller:2004tu,Giudice:2003jh,Davidson:2008bu}). In this paper, we focus on the {\it vanilla} leptogenesis in the context of MES model. Leptogenesis in the MES scheme has been discussed in \cite{Das:2019kmn,Das:2020vca} with keV scale sterile neutrino. As we will see in the next section, incorporating the bounds from oscillation experiments including the ones on active-sterile mixing would imply that $M_1 \gtrsim 10^{11}$ GeV with most of the points lying above $10^{12}$ GeV which is the unflavored regime of leptogenesis. Thus, the expression for the CP asymmetry guided by the decay of the lightest RHN in this model can be written as~\cite{Davidson:2008bu},
\begin{equation}
\epsilon_1 = \frac{1}{8 \pi v^2} \frac{1}{(M_D M_D^\dagger) _{11}}\sum_{j = 2,3} \rm{Im} \{(M_D M_D^\dagger) _{1j}^2\} f (x),
\end{equation}
where, the loop function can be expressed as $f(x)= \sqrt{x}\Big(1-(1+x)\text{ln}(\frac{1+x}{x})-\frac{1}{1-x}\Big)$ with $x = \frac{M_j^2}{M_1^2}$. For $x \gg 1$, {\it i.e.,} when a large hierarchy exists among the RHN mass states, one can simply write $f(x) \approx - \frac{3}{2 \sqrt{x}}$.  
After determining the lepton asymmetry $\epsilon_1$ using the above expression, the corresponding baryon asymmetry can be obtained through the electroweak sphaleron processes as~\cite{PhysRevD.49.6394,DOnofrio:2012phz},
\begin{equation}
Y_B = 1.27 \times 10^{-3} \epsilon_1 \eta(\tilde m_1).
\end{equation}
Here, the factor 
\be \tilde m_1 = (M_D M_D^\dagger)_{11} /M_1 \ee is a measure of the effective neutrino mass which contains the information on solar and atmospheric mass splittings. The nature of the wash out regime is also decided by the efficiency factor $\eta$, which is as~\cite{Davidson:2008bu,Rodejohann:2010zz},
 \be \label{eq:efficiency} \eta(\tilde{m_{1}})\approx 1/((8.25 \times 10^{-3} \text{eV})/\tilde{m_{1}} + (\tilde{m_{1}}/(2 \times 10^{-4} \text{eV})^{1.16}) . \ee
Departure from thermal equilibrium\footnote{which is the Sakharov's third condition \cite{Sakharov:1967dj} to be satisfied to have a baryon asymmetry.} can be estimated by comparing the interaction rate with the expansion rate of the Universe. At a very high temperatures $T \geq 10^{12}$ GeV, all charged lepton flavors are out of equilibrium, and hence all of them behave indistinguishably resulting in the {\it vanilla} leptogenesis scenario. The decay parameter which governs the competition between the decay rate and expansion rate of the Universe can be written as,
\begin{equation}
K = \frac{\Gamma_1}{H (T = M_1)} = \frac{(M_D M_D^\dagger)_{11}M_1}{8 \pi v^2} \frac{M_{\rm pl}}{1.66 \sqrt{g_*} M_1^2},
\end{equation}
 where $\Gamma_1$ is the decay rate of the lightest RHN, ${\nu_R}_1$, and $H(T = M_1)$ is the Hubble expansion rate at temperature $T = M_1$. The effective number of relativistic degrees of freedom is measured by the quantity $g_*$ which is 106.75 \cite{Bauer:2017qwy}. Depending on $K$, one can have an idea whether it is in agreement with the Sakharov's third condition or not. 

It is also instructive to examine the washout regime in the scenario of thermal leptogenesis, which relies on the parameter $\eta$ as evident in the Eq.~(\ref{eq:efficiency}). The efficiency factor is directly connected to $K$ through the parameter $\tilde{m}$, which again depends on the order of neutrino mass squared differences. In our analysis, the efficiency factor is obtained to be of the order of $10^{-4} \,-\, 10^{-3}$, which gives an insight of the amount of washout produced. This order of the washout strictly falls within the strong regime, which is also favored by the observed neutrino mass squared differences~\cite{Buchmuller:2004nz}.

\section{Numerical analysis and Results} \label{sec:analysis}

In this section, we discuss the numerical analysis performed and the results obtained in detail. We investigate the parameter space that allows successful thermal leptogenesis and at the same time satisfying the bounds from the 3+1 mixing data. An adequate amount of lepton asymmetry is essentially sourced by the complex Yukawa coupling ($y_\nu$) which governs the RHN decay to the SM lepton and the Higgs doublet. As discussed above, the presence of additional sterile states in the seesaw mechanism induces deviation of the neutrino mixing matrix from being unitary. Note that the canonical type-I seesaw also admits non-unitarity and is determined by the factor $M_D^2 /M_R^{2}$, which is very small ($\sim \mathcal{O}(10^{-20})$). However, for the extended seesaw mechanism which has an additional eV sterile neutrino, the non-unitarity is determined by the ratio $M_D^2/ M_S^2$, which can be $\sim \mathcal{O}(0.1)$. Assuming that such a sterile neutrino is responsible for the LSND-MiniBooNE anomalies, the results of the fit of short-baseline neutrino oscillation data for the 3+1 active-sterile neutrino mixing are given in table \ref{tab:input}~\cite{Dentler:2018sju,Aghanim:2018eyx,Aker:2019qfn,Gariazzo:2015rra}.

\begin{table}[t]
\begin{center}
\begin{tabular}{|c|c|}
\hline
& $3\sigma$ range\\
\hline
$\sin^2 \theta_{12}$ & $0.24 \rightarrow 0.377$ \\
$\sin^2 \theta_{13}$ & $0.02044 \rightarrow 0.02437$ \\
$\sin^2 \theta_{23}$ & $0.48 \rightarrow 0.599$ \\
$\Delta m^2_{21}/\text{eV}^2$ & $6.79\times10^{-5} \rightarrow 8.01\times10^{-5}$\\
$\Delta m^2_{31}/\text{eV}^2$ & $2.431\times10^{-3} \rightarrow 2.622\times10^{-3}$\\
$\Delta m^2_{41}/\text{eV}^2$ & $0.87\rightarrow 2.04$\\
$|V_{e4}|^2$ &  $0.012 \rightarrow 0.047$\\
 $|V_{\mu4}|^2$ &  $0.005\rightarrow 0.03$\\
 $|V_{\tau4}|^2$ & $< 0.16$\\
\hline
\end{tabular}
\caption{The $3\sigma$ ranges for the 3+1 neutrino oscillation parameters \cite{Dentler:2018sju,Aghanim:2018eyx,Aker:2019qfn,Gariazzo:2015rra} that are used in our analysis.}
\label{tab:input}
\end{center}
\end{table}

 To find out the parameter space that gives the correct baryon asymmetry, we first determined the values of the Dirac mass matrices, $M_D$ (which is just $y_\nu v/\sqrt{2}$), that satisfies all the low energy data with the help of the CI parametrization discussed in section \ref{sec:YukawaCI}. For this, we did a random scanning over all the neutrino oscillation parameters in their $3\sigma$ ranges which are given in table \ref{tab:input}. Care has been taken to abide by the hierarchy of the mass scales as $M_D< M_S<< M_R$, as is required for the MES model and thereby ensuring the validity of the seesaw approximations. This particular hierarchy among the mass scales not only ensures the light sterile neutrino to have a mass in the eV regime, but also facilitates in maintaining the active-sterile mixing strength which is complied by the experimental data. We have taken only those points for which $\mathcal{O}(M_D) \leq 0.1 \,\mathcal{O}(M_S)$. We have neglected the next-to-leading order corrections to the active neutrino mass matrix, which is of the order of $\frac{M_D^4}{M_S^2 M_R}$ \cite{Nath:2016mts}, as it is always $\leq 5 \times 10^{-5}$ in the parameter space that we have considered. In our scanning, we have chosen the following ranges for the entries of the corresponding mass matrices,
\begin{gather}\label{eq:ranges}
200 ~{\rm GeV} \le |M_{S}^{i1}| \leq 3000 ~{\rm GeV} \,\,\,,\,\,\, 
10^{8} ~{\rm GeV} \le M_{R}^{ii}\leq 10^{16} ~{\rm GeV} \,\,\, (i=1,2,3).
\end{gather}
We have chosen $M_R$ to be diagonal and real whereas for the $3\times 1$ matrix $M_S$, phases of the entries are varied in the range $0-2\pi$. The Dirac and Majorana phases that enter $U_\nu$ are also varied in the ranges $0-2\pi$ and $0-\pi$, respectively. These five phases act as the sources of CP violation (We have kept the orthogonal matrix $R$ to be real for simplicity with the angles varying in the range $0-2\pi$). Using these ranges and the $3\sigma$ oscillation parameters, we performed  a random scanning over $3\times 10^8$ data points to find out $M_D$ ( or $y_\nu$) and then we calculated the baryon asymmetry $Y_B$ using the expressions given in section \ref{sec:leptogenesis}\,. To show the impact of the inclusion of the non-unitary corrections due to active-sterile mixing on the baryon asymmetry, we have given a few benchmark points in table \ref{data}\,. Four different points are given corresponding to different masses of the lightest heavy RHN, $M_1$. The third and the fifth columns show the values of $Y_B$ calculated just from the unitary part and from the inclusion of non-unitary parts, respectively. The former is only an artificial case where we
put the $V^\dag V$ term as zero by hand and hence takes into account only the dimension-5 contribution to $Y_B$.  It can be seen that the inclusion of the $V^\dag V$ term (which is the dimension-6 contribution) has a considerable impact on $Y_B$ and thereby indicates the dependence of $Y_B$ on active-sterile mixing. 


\begin{table}[ht]
 $$
 \begin{array}{|c|c|c|c|c|c|}
 \hline M_1 (GeV) & m_{sterile} (eV) &  Y_B (\textrm{from unitary part}) & \epsilon_1 (\textrm{from unitary part}) & Y_B (\textrm{total})  & \epsilon_1 (\textrm{total})  \\
  
 \hline
 
  4.2624 \times 10^9 & 1.58356 & 2.07665\times 10^{-16} & 1.38397\times 10^{-9} & 7.91104\times 10^{-16} & 4.1033\times 10^{-10} \\
  
  \hline
 
4.32082 \times 10^{10} & 1.98397 & 6.23251\times 10^{-16} & 3.6963\times 10^{-8} & 2.45091\times 10^{-13}  & 3.80607\times 10^{-7} \\
   
 \hline

1.01752\times 10^{12} & 0.938872 & 2.93063\times 10^{-14} & 9.77027\times 10^{-8} & 9.32359\times 10^{-13} & 1.22387\times 10^{-6}  \\
   
 \hline
 
 1.00007\times 10^{13} & 1.16399 & 3.29334\times 10^{-11} & 0.000107207 & 1.29023\times 10^{-11} &  3.43336\times 10^{-6} \\
   
 \hline

 \end{array}
 $$
\caption{Baryon ($Y_B$) and CP ($\epsilon_1$) asymmetries with and without including the non-unitary corrections for different benchmark points. }\label{data}\end{table}

\begin{figure*}[h!]
\begin{center}
\includegraphics[scale=0.5]{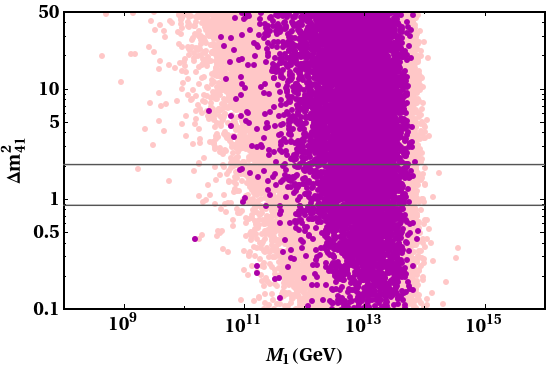}
\caption{Variation of $\Delta m_{41}^2$ with respect to the mass of lightest RHN. The region within the gray lines corresponds to the range of $\Delta m^2_{41} $ allowed by the experiments. The pink points do not satisfy the bounds on active sterile mixing but they satisfy the constraints from three neutrino masses and mixing. The magenta points satisfy the bounds on active-sterile mixing that are shown in table\,\,\ref{tab:input}\,.}
\label{MN1_deltm41sq}
\end{center}
\end{figure*}

In Fig.~\ref{MN1_deltm41sq}, we have shown the variation of $\Delta m_{41}^2$ with respect to the mass of the lightest RHN. Here, the region within the gray lines corresponds to the range of $\Delta m^2_{41} $ allowed by experiments,\,($ 0.87-2.04~\textrm{eV}^2$). The use of CI parametrization implies that the bounds from the three neutrino mixing are satisfied. The pink points do not satisfy the bounds on active sterile mixing whereas the magenta points satisfy these bounds that are shown in table \ref{tab:input}\,. Thus, one can note from this figure that the requirement of having an eV scale sterile neutrino itself imposes a lower bound on the value of $M_1$ as $\sim 10^{10}$ GeV, as can be seen from the region within the gray band. Once the bound on active-sterile mixing are incorporated, this lower bound on $M_1$ increases further to $\sim 10^{11}$ GeV. 

\begin{figure*}[h!]
\begin{center}
\includegraphics[scale=0.5]{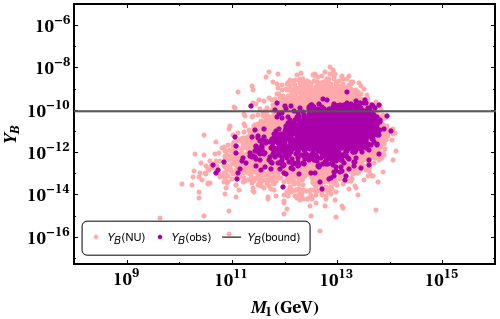}
\caption{Variation of $Y_B$ with respect to the mass of lightest RHN. No constraints are put on the active-sterile mixing for the pink points, but these points satisfy the constraints on three neutrino mixing and has an eV scale sterile neutrino with $\Delta m^2_{41} $ in the range $ 0.87-2.04~\textrm{eV}^2$. The magenta points correspond to the parameter space where the active-sterile mixing comply with the experimental data. The grey thick line corresponds to the observed baryon asymmetry of the universe.}
\label{YB_RHNmass}
\end{center}
\end{figure*}

 Fig.~\ref{YB_RHNmass} shows the variation of the baryon asymmetry yield with respect to the mass of the lightest RHN.  In this figure, no constraints are put on the active-sterile mixing for the pink points, but they satisfy the constraints on three neutrino mixing and has an eV scale sterile neutrino with $\Delta m^2_{41} $ in the range $ 0.87-2.04~\textrm{eV}^2$. The magenta points correspond to the parameter space where the active-sterile mixing satisfy the bounds from the experimental data, as given in table\,\,\ref{tab:input}\,. The gray thick line corresponds to the observed baryon asymmetry of the universe, $Y_B = (8.52 - 8.93)\times10^{-11}$. We have seen from Fig.~\ref{MN1_deltm41sq} that the model itself imposes a constraint on the mass of the lightest heavy RHN to be $M_1 \gtrsim  10^{11}$ GeV, once the bounds on the mass-squared difference and active-sterile mixing are incorporated. 
It can be seen from Fig.~\ref{YB_RHNmass} that for this model to account for the entire observed baryon asymmetry, $M_1$ has to be greater than $  10^{11}$ GeV (the region where the thick gray line overlaps with the pink points). The value of $M_1$ is restricted further once the bounds on active-sterile mixing are included as is shown by the magenta points (the region where the gray thick line overlaps with the magenta points). In fact, $M_1$  prefers to take values $\gtrsim 10^{12}$ GeV to give the correct baryon asymmetry. As mentioned earlier, this is the unflavored regime of leptogenesis where the flavors are indistinguishable. Note that this value of $M_1$ is higher than the Davidson-Ibarra bound of $10^8-10^9$ GeV for the canonical type-I seesaw model~\cite{Davidson:2002qv}.

 \begin{figure*}[t]
\begin{center}
\includegraphics[scale=0.45]{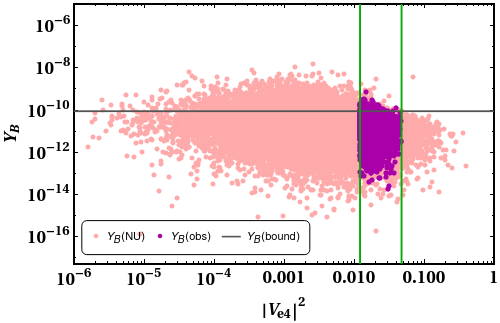}
\includegraphics[scale=0.45]{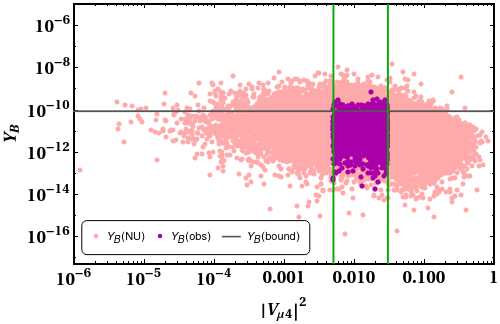}\\    
\includegraphics[scale=0.45]{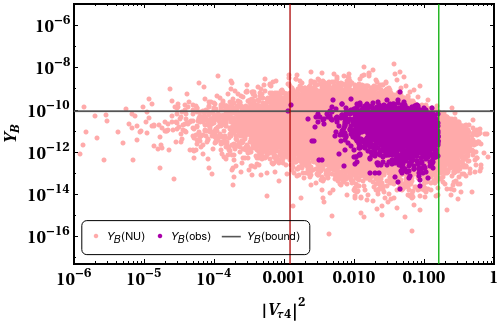}    
\caption{Variation of $Y_B$ with respect to the active-sterile mixing, $|V_{e4}|^2$, $|V_{\mu 4}|^2$ and $|V_{\tau 4}|^2$. The colour codes are the same as in Fig.~\ref{YB_RHNmass}. The green lines are used to indicate the upper and/or lower bound of the matrix elements $|V_{\alpha 4}|^2$ provided by the relevant experiments. The red line shows the lower bound on $|V_{\tau 4}|^{2}$ which we have obtained from our analysis.}
\label{YB_mixing}
\end{center}
\end{figure*}

In Fig.~\ref{YB_mixing}\,,  we show the variation of $Y_B$ with respect to the three active-sterile mixing, $|V_{e4}|^2$ (upper left panel), $|V_{\mu 4}|^2$ (upper right panel) and $|V_{\tau 4}|^2$ (lower panel). The colour codes are the same as in Fig.~\ref{YB_RHNmass}. The green lines are used to indicate the upper and/or lower bound of the mixing parameters from global analysis given in table\,\,\ref{tab:input}\,. In our scanning, these parameters take values from $10^{-6}$ to 1 and the figures show that even for very small values of these mixing parameters, successful baryogenesis can be obtained. The magenta points in the upper panels show that the constraint on $Y_B$ is satisfied in the current allowed ranges of $|V_{e4}^2|$ and $|V_{\mu 4}^2|$. In the case of $|V_{\tau 4}|^2$, there exists no lower bound  from the current experimental data and there is only an upper bound of $0.16$. However, the model gives a lower bound on $|V_{\tau 4}|^2$ as $\sim 0.001$. This lower bound 
is coming due to the the large values of $|V_{e4}^2|$ and $|V_{\mu 4}^2|$ in the region allowed by the experiments. We can see from the figures that this lower bound on  $|V_{\tau 4}|^2$ is not there once we let go off the bounds on the other two mixing elements, as is shown by the pink points.

 \begin{figure*}[h]
\begin{center}
\includegraphics[scale=0.45]{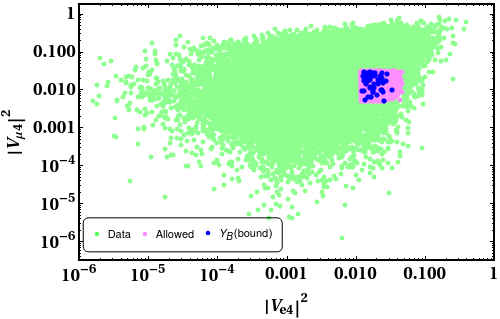}
\includegraphics[scale=0.45]{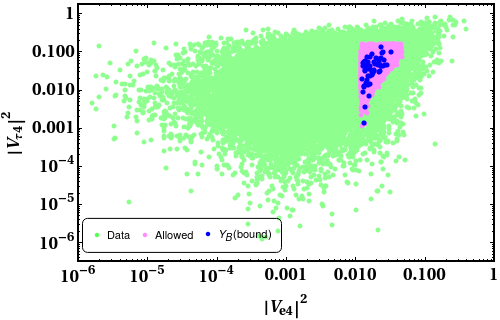}    
\includegraphics[scale=0.45]{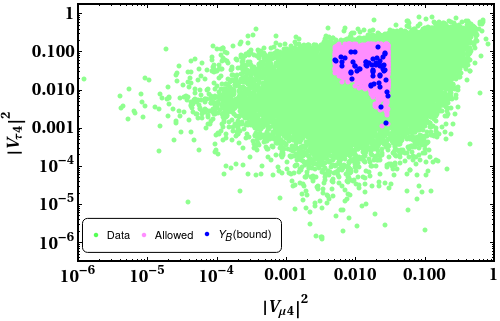}    
\caption{Correlations among the active-sterile mixing elements by taking various constrains into account. The green region correspond to the parameter space that satisfies the bounds from $3$ neutrino mixing and there exists a light sterile neutrino of mass in the range (0.87 - 2.04) eV$^{2}$. The pink points correspond to the parameter space which satisfies the bounds on the active-sterile mixing as shown in table \ref{tab:input}. The blue points indicate the regions that give the correct values for the observed baryon asymmetry of the universe, that is, in the range $Y_B = (8.52 - 8.93)\times10^{-11}$.}
\label{mixing}
\end{center}
\end{figure*}

In Fig.~\ref{mixing}\,, we present the correlations among the active-sterile mixing elements obtained from the imposition of the various constraints on the total parameter space. As seen earlier, there can be a large region of parameter space which obey all the 3-neutrino oscillation data as depicted by the green region in this figure. However, it is to be noted  here that this is not the final parameter space we are looking for. The pink points in this figure correspond to the parameter space which satisfies the bounds on the active-sterile mixing as shown in table\,\,\ref{tab:input} and the blue points indicate the regions that give the correct values for the observed baryon asymmetry, that is, in the range $Y_B = (8.52 - 8.93)\times10^{-11}$. We can clearly see the existence of a lower bound on $|V_{\tau 4}|^2$ as $\sim 0.001$ in this model, once the bounds on $|V_{e 4}|^2$ and $|V_{\mu 4}|^2$ are incorporated, as was seen in the previous figure. In addition, there is no real correlation between $|V_{e 4}|^2$ and $|V_{\mu 4}|^2$ in the parameter space allowed by the experiments (pink region of the upper left panel in Fig.~\ref{mixing}). On the other hand, relatively higher values of $|V_{\tau 4}|^2$ are preferred for higher values of $|V_{e 4}|^2$ (pink region of the upper right panel in Fig.~\ref{mixing}) and lower values of $|V_{\mu 4}|^2$ (pink region of the lower panel in Fig.~\ref{mixing}). Values of $|V_{\tau 4}|^2$ in the range $\sim 0.001-0.16$ are allowed corresponding to the lower limit on $|V_{e 4}|^2$ and upper limit on $|V_{\mu 4}|^2$ respectively. As can be seen from the blue points, successful baryogenesis is possible in most of the regions allowed by the experiments for the eV scale sterile neutrino.

\section{Conclusions}\label{sec:conclusion}

 In this work, we have studied the effects of an eV scale sterile neutrino on leptogenesis in the context of the minimal extended seesaw model (MES). This model contains a light sterile neutrino in addition to the three heavy Majorana right handed neutrinos. Here, the active-light sterile mixing act as non-unitary parameters introducing considerable deviation of the PMNS matrix from being unitary. We noted that the constraints on the active-sterile mixing coming from global analysis of the data from various short baseline experiments impose an upper bound on the lightest heavy neutrino mass scale ($M_1$) as $\gtrsim 10^{11}$ GeV in this model. This is an artifact of the modified expression for $M_D$ as well as the requirement of having an eV scale sterile neutrino. In addition, we also found that there exists a lower bound of $\sim 0.001$, on the active-sterile mixing element $|V_{\tau 4}|^2$  once the bounds on $|V_{e 4}|^2$ and $|V_{\mu 4}|^2$ are incorporated. This is an important prediction from the model since the analysis of current data only gives an upper bound of 0.16 on $|V_{\tau 4}|^2$. 

Coming to the implications for leptogenesis, we studied the standard {\it vanilla} leptogenesis where the out of equilibrium decay of the heavy right handed Majorana neutrinos in the early universe can generate a lepton asymmetry, which in turn can be converted into a baryon asymmetry by the non-perturbative sphaleron processes. We used the Casas-Ibarra parametrization of the Dirac mass term for active neutrinos to facilitate our numerical analysis.  We found that the incorporation of the bounds on active-sterile mixing raises the lower bound on $M_1$ to be $\gtrsim 10^{12}$ GeV, and thereby makes the flavor effects on leptogenesis insignificant in this parameter space. Thus, we noted that even though it might look as if the light sterile neutrino plays no role in leptogenesis, the bounds on active-sterile mixing actually shrinks the parameter space where successful explanation of the observed baryon asymmetry of the universe is possible. We have also studied the correlations of $Y_B$ to the active-sterile mixing parameters and noted that successful baryogenesis is possible in most of the regions allowed by the experiments. In summary, the non-unitary effects in the MES model can give rise to interesting consequences for leptogenesis.

\section*{Acknowledgement}
S.G. acknowledges  the J.C Bose Fellowship  (JCB/2020/000011)  of Science and  Engineering Research Board  of Department of Science and Technology, Government of India. AM would like to acknowledge the financial support provided by SERB-DST, Govt. of India through the project EMR/2017/001434.

\bibliographystyle{utphys}
\bibliography{mes_lepto}
\end{document}